\documentclass[12pt,letter]{article}

\usepackage{graphicx, epsfig, color}
\def\to{\rightarrow}

\def\bi{\begin{itemize}}
\def\ei{\end{itemize}}
\def\te{\tilde e}

\def\tu{\tilde u}
\def\sps1ap{SPS1a$^\prime$}
\def\c1p{C1$^\prime$}

\def\tf{\tilde f}
\def\td{\tilde d}

\def\tst{\tilde t}

\def\tnu{\tilde\nu}
\def\tell{\tilde\ell}
\def\tq{\tilde q}

\def\alt{\stackrel{<}{\sim}}
\def\agt{\stackrel{>}{\sim}}
\def\be{\begin{equation}}  
\def\ee{\end{equation}}  
\def\bea{\begin{eqnarray}}  
\def\eea{\end{eqnarray}}  
\def\beas{\begin{eqnarray*}}  
\def\eeas{\end{eqnarray*}}  
\newcommand\prd[3]{{\it Phys.\ Rev.\ }{\bf D #1} (#2) #3}

\newcommand\plb[3]{{\it Phys.\ Lett.\ }{\bf B #1} (#2) #3}

\begin{document}
\begin{titlepage}
\begin{flushright}
UH-511-1225-13
\end{flushright}

\vspace{0.5cm}
\begin{center}
{\Large \bf Naturalness implies intra-generational degeneracy\\
for decoupled squarks and sleptons
}\\ 
\vspace{1.2cm} \renewcommand{\thefootnote}{\fnsymbol{footnote}}
{\large Howard Baer$^1$\footnote[1]{Email: baer@nhn.ou.edu },
Vernon Barger$^2$\footnote[2]{Email: barger@pheno.wisc.edu },
Maren Padeffke-Kirkland$^1$\footnote[3]{Email: m.padeffke@ou.edu } 
Xerxes Tata$^3$\footnote[4]{Email: tata@phys.hawaii.edu } 
}\\ 
\vspace{1.2cm} \renewcommand{\thefootnote}{\arabic{footnote}}
{\it 
$^1$Dept. of Physics and Astronomy,
University of Oklahoma, Norman, OK 73019, USA \\
}
{\it 
$^2$Dept. of Physics,
University of Wisconsin, Madison, WI 53706, USA \\
}
{\it 
$^3$Dept. of Physics and Astronomy,
University of Hawaii, Honolulu, HI 96822, USA \\
}

\end{center}

\vspace{0.5cm}
\begin{abstract}
\noindent 
The SUSY flavor, CP, gravitino and proton-decay problems are all solved
to varying degrees by a decoupling solution wherein first/second
generation matter scalars would exist in the multi-TeV regime.  Recent
models of natural SUSY presumably allow for a co-existence of
naturalness with the decoupling solution. We show that:
if sfermions are heavier than $\sim 10$~TeV, then a small
first/second generation contribution to electroweak fine-tuning (EWFT)
requires a rather high degree of intra-generational degeneracy of either 
1. (separately) squarks and sleptons, 
2. (separately) left- and right-type sfermions, 
3. members of $SU(5)$ multiplets, or 
4. all members of a single generation as in $SO(10)$. 
These (partial) degeneracy patterns required by naturalness hint at the 
necessity of an organizing principle, and highlight the limitations of 
models such as the pMSSM in the case of 
decoupled first/second generation scalars.

\vspace*{0.8cm}

\end{abstract}

\end{titlepage}

\section{Introduction}

Weak scale supersymmetry provides a solution to the notorious gauge
hierarchy problem by ensuring the cancelation of quadratic divergences
endemic to scalar fields which are otherwise unprotected by a
symmetry\cite{wss}. While realistic and natural SUSY models of particle physics
can be constructed in accord with all experimental constraints,
especially those arising from recent LHC searches, they are subject to a
host of open questions\cite{dk}. Included amongst these are
\begin{itemize}
\item the SUSY flavor problem\cite{masiero}, wherein unfettered flavor-mixing soft terms lead to
{\it e.g.} large $K-\bar{K}$ mass difference and anomalous contributions to flavor-changing decays
such as $b\to s\gamma$ and $\mu\to e\gamma $,
\item the SUSY $CP$ problem\cite{masiero}, in which unfettered $CP$ violating phases 
lead to large contributions to electron and various atomic EDMs, 
\item the SUSY gravitino problem\cite{weinberg}, wherein thermally produced gravitinos in the early universe
may decay after BBN, thus destoying the successful prediction of light element abundances 
created in the early universe, and
\item the SUSY proton decay problem\cite{mp}, wherein even in $R$-parity conserving GUT theories, the
proton is expected to decay earlier than recent bounds from experimental searches.
\end{itemize}
While there exist particular solutions to each of these problems ({\it
e.g.}  degeneracy\cite{dg} or alignment\cite{seiberg} for the flavor
problem, small phases for $CP$ problem, low $T_R$ for gravitino
problem\cite{linde}, cancellations for proton decay\cite{nath}), there
is one solution which potentially tames all four: decoupling of squarks
and sleptons\cite{dine,nelson,bagger}.\footnote{In the case of the
gravitino problem, we tacitly assume here gravity-mediation of SUSY
breaking, wherein the
scalar mass parameters as well as the gravitino mass $m_{3/2}$ arise from a 
common source of SUSY breaking in a hidden sector. In this case,
the scalar mass parameters all have magnitudes comparable to $\sim m_{3/2}$.}
For the decoupling
solution, squark and slepton masses $\agt$ a few TeV is sufficient for
the SUSY $CP$ problem while $m_{3/2}\agt 5$ TeV allows for gravitino
decay before the onset of BBN. For the SUSY flavor problem, then
first/second generation scalars ought to have mass $\agt 5-100$ TeV
depending on which process is examined, how large of flavor-violating soft terms are allowed
and possible GUT relations amongst GUT scale soft terms\cite{moroi}. 
For proton decay, again multi-TeV matter scalars seem sufficient to suppress decay 
rates depending on other GUT scale parameters\cite{ccn,hisano}.

Naively, the decoupling solution seems in conflict with notions of SUSY
naturalness, wherein sparticles are expected at or around the weak
scale\cite{dg} typified by the recently discovered Higgs mass
$m_h=125.5\pm 0.5$ GeV\cite{atlas_h,cms_h}.  To move beyond 
this, we require the necessary (although not sufficient) condition for
naturalnesss, quantified by the measure of electroweak fine-tuning (EWFT)
which requires that there be 
{\it no large cancellations within the weak scale contributions to $m_Z$ or 
to $m_h$}\cite{ltr,rns,perel,ccn,comp}.  

Recall that minimization of the one-loop effective potential $V_{\rm tree} + \Delta V$ leads to
the well-known relation
\be \frac{M_Z^2}{2} = \frac{m_{H_d}^2 + \Sigma_d^d - 
(m_{H_u}^2+\Sigma_u^u)\tan^2\beta}{\tan^2\beta -1} -\mu^2 \;,
\label{eq:mz}
\ee 
where $\Sigma_u^u$ and $\Sigma_d^d$ are radiative corrections that
arise from the derivatives of $\Delta V$ evaluated at the potential minimum.
Noting that all entries in Eq.~(\ref{eq:mz}) are defined at the weak scale, 
the {\it  electroweak fine-tuning measure} 
\be 
\Delta_{EW} \equiv max_i \left|C_i\right|/(m_Z^2/2)\;, 
\ee 
may be constructed, where $C_{H_d}=m_{H_d}^2/(\tan^2\beta -1)$,
$C_{H_u}=-m_{H_u}^2\tan^2\beta /(\tan^2\beta -1)$ and $C_\mu =-\mu^2$.
Also, $C_{\Sigma_u^u(k)} =-\Sigma_u^u(k)\tan^2\beta /(\tan^2\beta -1)$
and $C_{\Sigma_d^d(k)}=\Sigma_d^d(k)/(\tan^2\beta -1)$, where $k$ labels
the various loop contributions included in Eq.~(\ref{eq:mz}).  Expressions
for the $\Sigma_u^u$ and $\Sigma_d^d$ are given in the Appendix of the
second paper of Ref. \cite{rns}.  The contributions from $\Sigma_u^u(k)$
are almost always much more important than the $\Sigma_d^d(k)$ since the
$\Sigma_d^d(k)$ are suppressed by the factor $1/\tan^2\beta$.
Typically, the dominant radiative corrections to Eq.~(\ref{eq:mz}) come
from the top-squark contributions $\Sigma_u^u(\tst_{1,2})$.  By adopting
a large value of the weak scale trilinear soft term $A_t$, then each of
$\Sigma_u^u(\tst_{1})$ and $\Sigma_u^u(\tst_{2})$ can be minimized
whilst lifting up $m_h$ into the 125 GeV regime\cite{ltr}.

For first/second generation sfermions, neglecting the small Yukawa
couplings, we find the contributions 
\bea \Sigma_{u,d}^{u,d}(\tf_{L,R}) = \mp
\frac{c_{col}}{16\pi^2}F(m_{\tf_{L,R}}^2)\left(-4g_Z^2(T_3-Q_{em}x_W) \right),
\label{eq:sigud}
\eea 
where $T_3$ is the weak isospin, $Q_{em}$ is the electric charge
assignment (taking care to flip the sign of $Q_{em}$ for $R$-sfermions),
$c_{col}=1 (3)$ for color singlet (triplet) states, 
$x_W\equiv\sin^2\theta_W$ and where 
\be
F(m^2)= m^2\left(\log\frac{m^2}{Q^2}-1\right) .  
\ee 
We adopt an optimized scale choice 
$Q^2 = m_{\rm SUSY}^2 \equiv
m_{\tst_1}m_{\tst_2}$.\footnote{The optimized scale choice is chosen to
minimize the log contributions to $\Sigma_u^u(\tst_{1,2})$ which occur
to all orders in perturbation theory.}  The explicit first generation
squark contributions to $\Sigma_u^u$ (neglecting the tiny Yukawa
couplings) are given by
\bea
\Sigma_u^u (\tu_L) &=& \frac{3}{16\pi^2}F(m_{\tu_L}^2)\left(-4g_Z^2
(\frac{1}{2}-\frac{2}{3} x_W)\right)\nonumber \\
\Sigma_u^u (\tu_R) &=& \frac{3}{16\pi^2}F(m_{\tu_R}^2)\left(-4g_Z^2 (\frac{2}{3}x_W)\right)\\
\Sigma_u^u (\td_L) &=& \frac{3}{16\pi^2}F(m_{\td_L}^2)\left(-4g_Z^2
(-\frac{1}{2}+\frac{1}{3}x_W)\right)\nonumber \\
\Sigma_u^u (\td_R) &=& \frac{3}{16\pi^2}F(m_{\td_R}^2)\left(-4g_Z^2
(-\frac{1}{3}x_W)\right) . \nonumber
\label{eq:Ssquarks}
\eea
These contributions, arising from electroweak $D$-term contributions to
masses, are frequently neglected since the various
contributions cancel amongst themselves in the limit of mass degeneracy
due to the fact that weak isospins and electric charges (or weak
hypercharges) sum to zero in each generation.  However, if squark and
slepton masses are in the multi-TeV regime but are {\it non-degenerate}
within each generation, then the contributions may be large and
non-cancelling.  In this case, they may render a theory which is
otherwise considered to be natural, in fact, unnatural.

The first generation slepton contributions to $\Sigma_u^u$ are given by
\bea
\Sigma_u^u (\te_L) &=& \frac{1}{16\pi^2}F(m_{\te_L}^2)\left(-4g_Z^2
(-\frac{1}{2}+x_W)\right) \nonumber \\
\Sigma_u^u (\te_R) &=& \frac{1}{16\pi^2}F(m_{\te_R}^2)\left(-4g_Z^2 (-x_W)\right)\\
\Sigma_u^u (\tnu_L) &=& \frac{1}{16\pi^2}F(m_{\tnu_{eL}}^2)\left(-4g_Z^2
(\frac{1}{2})\right) ; \nonumber
\label{eq:Ssleptons}
\eea
these may also be large for large $m_{\tell}^2$ although again they 
cancel amongst themselves in the limit of slepton mass degeneracy.

Our goal in this Brief Report is to examine the case where the scalar
masses are large, as suggested by the decoupling solution, but where the
masses are not necessarily degenerate.  In models such as radiatively
driven natural SUSY\cite{rns}-- where $m_{H_u}^2$, $\mu^2$ and
$\Sigma_u^u(\tst_{1,2})$ are all $\sim 100-200$ GeV -- then for
non-degenerate first generation squarks and sleptons, the
$\Sigma_u^u(\tq_i)$ and $\Sigma_u^u(\tell_i)$ may be the dominant
radiative corrections: and if they are sufficiently large, then large
cancellations will be needed amongst independent contributions to yield
a value of $m_Z$ of just $\sim 91.2$ GeV: {\it i.e.} the model will
become highly electroweak fine-tuned.  Alternatively, requiring
electroweak naturalness (low $\Delta_{EW}\alt 30$) will require a rather
high degree of intra-generational degeneracy amongst decoupled
matter scalars.

\section{Results}

To a very good approximation, the masses of first and second 
generation sfermions (whose Yukawa couplings can be neglected) are given by
\be
m_{\tf_{i}}^2 = m_{F_i}^2 + m_{f_i}^2 + M_Z^2\cos 2\beta
\left(T_3-Q_{em}\sin^2 \theta_W\right)\simeq m_{F_i}^2 ,
\ee
where $m_{F_i}^2$ is the corresponding weak scale soft-SUSY breaking
parameter, and the sign of $Q_{em}$ is flipped for $R$-sfermions as
described just below Eq.~(\ref{eq:sigud}). The latter approximate equality
holds in the limit of large soft masses (decoupling), where $D$-term contributions
are negligible.

In the limit of negligible hypercharge $D$-terms and $m_{f_i}^2$, then
the elements of each squark and slepton doublet are essentially mass
degenerate; in this case, the weak isospin contributions to
Eq.~(\ref{eq:sigud}) cancel out, and one is only left with the
possibility of non-cancelling terms which are proportional to electric
charge. The summed charge contributions (multiplied by $c_{col}$) of
each multiplet are then $Q(Q_1)=+1$, $Q(U_1)=-2$, $Q(D_1)=+1$,
$Q(L_1)=-1$ and $Q(E_1)=+1$. To achieve further cancellation, one
may then cancel the $Q(U_1)$ against any two of $Q(Q_1)$, $Q(D_1)$ and
$Q(E_1)$. The remaining term may cancel against $Q(L_1)$. 
Thus, the
possible cancellations break down into four possibilities:
\begin{enumerate}
\item separate squark and slepton degeneracy: $m_{U_1}=m_{Q_1}=m_{D_1}$ and $m_{L_1}=m_{E_1}$,
\item separate right- and left- degeneracy: $m_{U_1}=m_{D_1}=m_{E_1}$ and $m_{L_1}=m_{Q_1}$,
\item $SU(5)$ degeneracy: $m_{U_1}=m_{Q_1}=m_{E_1}\equiv m_{{10}_1}$ and 
$m_{L_1}=m_{D_1}\equiv m_{5_1}$ and
\item $SO(10)$ degeneracy: $m_{U_1}=m_{Q_1}=m_{E_1}= m_{L_1}=m_{E_1}\equiv m_{{16}_1}$.
\end{enumerate} 
We assume that the gaugino masses are small enough
so that splittings caused by the renormalization of the mass parameters
between the GUT scale and the SUSY scale is negligible so that these
relations may equally be taken to be valid at the GUT scale.
Any major deviation from the first three of these patterns 
(which implies a deviation to the fourth $SO(10)$ pattern) 
can lead to unnaturalness in models with decoupled scalars. 
In models such as the phenomenological MSSM, or pMSSM, 
where $m_{U_1}$, $m_{Q_1}$, $m_{E_1}$, $m_{L_1}$ and $m_{E_1}$ are all taken as independent, 
a decoupling solution to the SUSY flavor, $CP$, gravitino and proton-decay problems
would likely be unnatural.

In this connection, it is worth mentioning that $D$-term contributions
associated with a reduction of rank when a GUT group is spontaneously
broken to the SM gauge symmetry can lead to intra-generational
splittings\cite{dterms}.  Assuming that weak hypercharge $D$-terms are
negligible, the splitting of the MSSM sfermions can be parametrized in
terms of the $vevs$ of the $D$-terms associated with $U(1)_X$ and
$U(1)_S$ (in the notation of the last paper of Ref. \cite{dterms}). The
$SU(5)$ splitting pattern 3. is automatically realized for arbitrary
values of $D_X$ and $D_S$, while patterns 1. and 2. do not appear to
emerge from the GUT framework.

\begin{figure}[tbp]
\includegraphics[height=0.4\textheight]{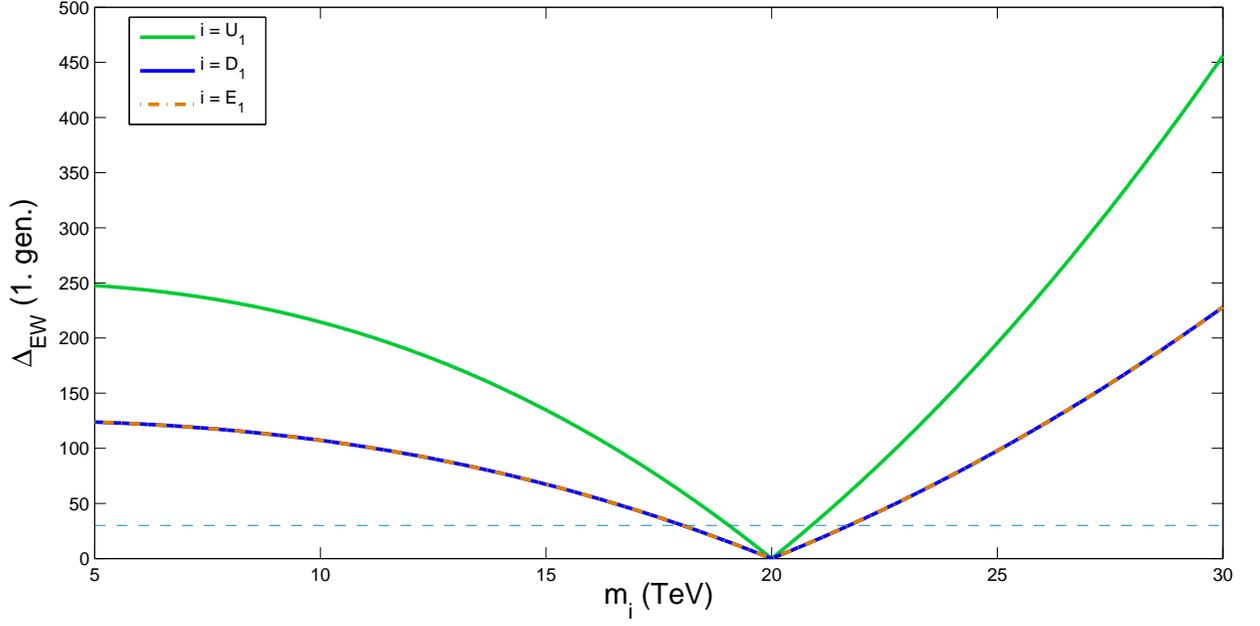}
\caption{Contribution to $\Delta_{EW}$ from first generation 
squarks and sleptons where all scalar soft masses are set 
to $20$ TeV except $m_{U_1}$ (green) or $m_{D_1}$ (blue) or $m_{E_1}$ (orange-dashed) 
with $m_{\rm SUSY}=2.5$ TeV and $\tan\beta =10$.
\label{fig:DEW1}}
\end{figure}

To illustrate the growth of $\Delta_{EW}$ for {\it ad hoc} sfermion
masses, in Fig. \ref{fig:DEW1} we plot as the green curve the summed
contribution to $\Delta_{EW}$ from first generation matter scalars by
taking all soft masses $m_{F_i}=20$ TeV except $m_{U_1}$ which varies
from 5-30~TeV.  The summed $\Sigma_u^u({\tilde f}_1)$ contributions to
$\Delta_{EW}$ for $m_{U_1}=5$ TeV begin at $\sim 250$ and slowly
decrease with increasing $m_{U_1}$.  The summed contributions reach zero
at $m_{U_1}=20$ TeV where complete cancellation amongst the various
squark/slepton contributions to $\Delta_{EW}$ is achieved.  A nominal
value of low EWFT adopted in Ref. \cite{rns} is 30: higher values of
$\Delta_{EW}$ require worse than $\Delta_{EW}^{-1}=3\%$ EWFT.  We see
from the plot that for $\Delta_{EW}<30$, then $m_{U_1}\sim 19-21$ TeV,
{\it i.e.} a rather high degree of degeneracy of $m_{U_1}$ in one of the
above four patterns is required by naturalness.

In Fig. \ref{fig:DEW1}, we also plot as the blue curve (with red dashes
lying atop) $\Delta_{EW}$ for all scalar soft masses $=20$ TeV except now varying
$m_{D_1}$. The contributions to $\Delta_{EW}$ are much reduced due to
the lower $d$-squark charge, but are still significant: in this case,
$m_{D_1}\sim 18-22$ TeV is required for $\Delta_{EW}<30$.  We also show
as the dashed red curve the contribution to $\Delta_{EW}$ from first
generation scalars where we take soft masses $=20$ TeV but now vary
$m_{E_1}$.  The curve lies exactly atop the varying $m_{D_1}$ curve
since the color factor of 3 in Eq.~(\ref{eq:Ssquarks}) exactly compensates
the increased electric charge by a factor three in
Eq.~(\ref{eq:Ssleptons}).  Thus, for $m_{F_1}=20$ TeV, then $m_{E_1}\sim
18-22$ TeV is required to allow for electroweak naturalness. Requiring $\Delta_{EW}$
as low as 10, as can occur in radiatively-driven natural
SUSY\cite{rns,comp}, requires even tighter degeneracy.

\begin{figure}[tbp]
\includegraphics[height=0.4\textheight]{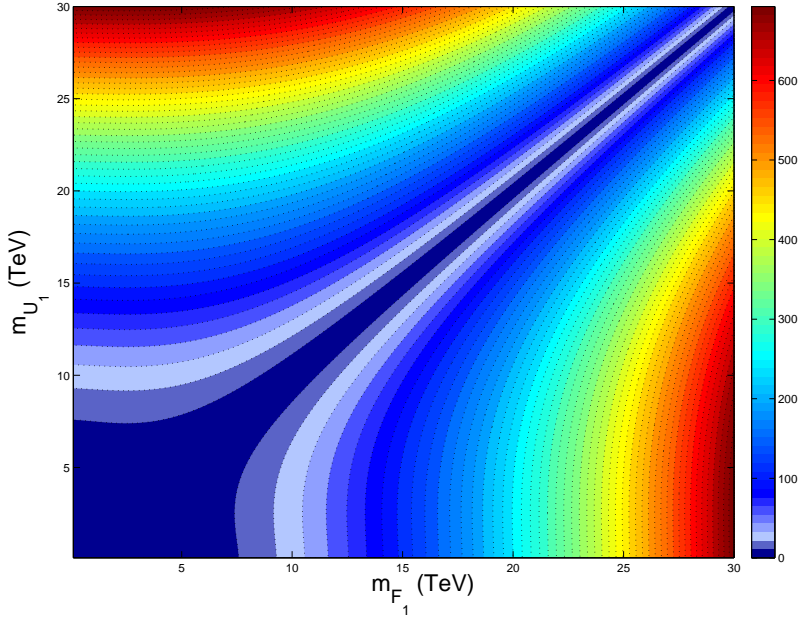}
\caption{Plot of contours of $\Delta_{EW}(\tf_1 )$ 
(summed over just first generation sfermions) in 
the $m_{U_1}$ vs. $m_{F_1}$ 
plane with $m_{\rm SUSY}=2.5$ TeV and $\tan\beta =10$.
\label{fig:mQvsmU}}
\end{figure}

Adopting a variant on the degenerate $SO(10)$ case with all sfermions but the
$\tu_R$ squark having the same mass, 
we plot in Fig. \ref{fig:mQvsmU} color-coded
regions of first generation squark contributions to $\Delta_{EW}$ in the
$m_{U_1}$ vs. $m_{F_1}$ plane, where $m_{F_a}$ stands for the common
sfermion mass other than $m_{U_1}$.  The regions in between the lightest 
grey bands (which have  $27 < \Delta_{EW}< 37$) would mark the rough boundary
of the natural region. From
the plot, we see that if weak scale soft squark masses are below $\sim
10$ TeV, then the $\Sigma_u^u(\tf_i)$ are all relatively small, and
there is no naturalness constraint on non-degenerate sfermion masses. As
one moves to much higher sfermion masses in the $\agt 10-15$ TeV regime,
then the sfermion soft masses within each generation are required to be
increasingly degenerate in order to allow for EW naturalness.

Similarly, we can show contributions to $\Delta_{EW}$ from first generation sleptons in
the $m_{L_1}$ vs. $m_{F_1}$ mass plane.  The various regions have
qualitatively similar shapes (but different widths, reflecting the
different coefficient $Q(L_1)$ that enters in the calculation) to
Fig. \ref{fig:mQvsmU} with the replacements $m_{U_1}\to m_{L_1}$: a high
degree of left-slepton mass degeneracy with another multiplet is
required by naturalness once slepton masses reach above about $10-15$
TeV.

\section{Conclusions:} 

The SUSY flavor, CP, gravitino and proton-decay problems are all solved
to varying degrees by a decoupling solution wherein first/second
generation matter scalars would exist in the multi-TeV regime.  In this
case, where matter scalar masses exist beyond the $\sim 10-15$ TeV level,
then intra-generation degeneracy following one of several patterns
appears to be necessary for electroweak naturalness, {\it i.e.} $\Delta_{EW}\alt 10-30$. 
Such degeneracy is not necessarily expected in generic SUSY models 
such as the pMSSM unless there is a protective
symmetry: for instance, $SU(5)$ or $SO(10)$ GUT symmetry provides
the required degeneracy provided additional contributions 
(such as running gauge contributions) are not very large.
Our results seem to hint at the existence of an additional 
organizing principle if a decoupling solution
(with sfermions heavier than $\sim 10$~TeV) to the SUSY flavor, CP,
gravitino and proton-decay problems is invoked along with
electroweak naturalness.
This could well be a Grand Unification symmetry,
in accord with recent calculations of flavor changing
contributions to $\Delta m_K$ where $SO(10)$ mass relations also 
contribute to suppress flavor violation\cite{moroi}.

\section*{Acknowledgments}

We thank an anonymous referee for very useful criticism.
This work was supported in part by the US Department of Energy, Office of High
Energy Physics.

%

%
\end{document}